\documentclass{SciPost}

\hypersetup{
    colorlinks,
    linkcolor={red!50!black},
    citecolor={blue!50!black},
    urlcolor={blue!80!black}
}

\usepackage{lmodern}
\usepackage[bitstream-charter]{mathdesign}
\usepackage{mathtools}
\usepackage{booktabs,tabularx}
\usepackage{microtype}
\usepackage{amsfonts}
\colorlet{scipostblue}{scipostdeepblue}
\DeclareSymbolFont{usualmathcal}{OMS}{cmsy}{m}{n}
\DeclareSymbolFontAlphabet{\mathcal}{usualmathcal}

\fancypagestyle{SPstyle}{
\fancyhf{}
\lhead{\colorbox{scipostblue}{\bf \color{white} ~SciPost Physics Proceedings }}
\rhead{{\bf \color{scipostdeepblue} ~Submission }}

\fancyfoot[C]{\textbf{\thepage}}
}
\newcommand{\dd}{\mathrm{d}}
\newcommand{\id}{\mathbb{I}}
\newcommand{\im}{\operatorname{Im}}
\newcommand{\Linfty}{\ensuremath{L_\infty}}
\newcommand{\divA}{\partial A}
\newcommand{\divE}{\partial E}
\newcommand{\divB}{\partial B}
\newcommand{\Boxinv}{\Box^{-1}}
\begin{document}
\pagestyle{SPstyle}

\begin{center}{\Large\textbf{\color{scipostdeepblue}{
Geometry of Gauge-invariant Quantum Fields
}}}\end{center}

\begin{center}\textbf{
Andrea Quadri\textsuperscript{1$\star$}
}\end{center}

\begin{center}
{\bf 1} INFN, Sezione di Milano, Milano, Italy
\\[\baselineskip]
% TODO: confermare l'indirizzo email prima della sottomissione.
$\star$ \href{mailto:andrea.quadri@mi.infn.it}{\small andrea.quadri@mi.infn.it}
\end{center}

\definecolor{palegray}{gray}{0.95}
% The logo is optional for compilation and can be added without changing
% the source. The full-width conference box is used while it is absent.
\begin{center}
\IfFileExists{logo_ICGTMP_nobckgrnd.png}{%
\colorbox{palegray}{%
\begin{tabular}{ll}
\hspace{-11pt}\begin{minipage}{0.18\textwidth}
\includegraphics[width=28mm]{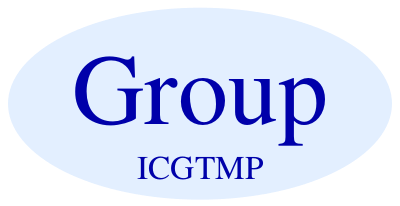}
\end{minipage}&
\begin{minipage}{0.75\textwidth}
\begin{center}
{\it 36th International Colloquium on Group Theoretical Methods in Physics}\\
{\it Valladolid, 13-17 July 2026}\\
\doi{10.21468/SciPostPhysProc.?}
\end{center}
\end{minipage}
\end{tabular}}%
}{%
\colorbox{palegray}{%
\begin{minipage}{0.94\textwidth}
\begin{center}
{\it 36th International Colloquium on Group Theoretical Methods in Physics}\\
{\it Valladolid, 13-17 July 2026}\\
\doi{10.21468/SciPostPhysProc.?}
\end{center}
\end{minipage}}%
}
\end{center}

\section*{\color{scipostdeepblue}{Abstract}}
\textbf{\boldmath{%
Gauge-invariant representatives of massless gauge fields generally
involve inverse differential operators. We trace this non-locality to the
homotopy contracting the free gauge complex and replace it, at field level,
by a local map using an independent scalar and a differential constraint.
We discuss the BRST completion needed to preserve the physical content and the
relation to the St\"uckelberg theory.
We finally comment on the extension of this approach to non-Abelian gauge groups via the extension to interactions through \Linfty\ homotopy transfer.
}}

\vspace{\baselineskip}
% Keep the publisher's placeholder block verbatim; adjust only the local
% table dimensions so that its two minipages fit inside the outer box.
\begingroup
\setlength{\textwidth}{0.975\linewidth}
\setlength{\tabcolsep}{0pt}
\setlength{\fboxsep}{2pt}
%%%%%%%%%% BLOCK: Copyright information
% This block will be filled during the proof stage, and finilized just before publication.
% It exists here only as a placeholder, and should not be modified by authors.
\noindent\textcolor{white!90!black}{%
\fbox{\parbox{0.975\linewidth}{%
\textcolor{white!40!black}{\begin{tabular}{lr}%
  \begin{minipage}{0.6\textwidth}%
    {\small Copyright attribution to authors. \newline
    This work is a submission to SciPost Phys. Proc. \newline
    License information to appear upon publication. \newline
    Publication information to appear upon publication.}
  \end{minipage} & \begin{minipage}{0.4\textwidth}
    {\small Received Date \newline Accepted Date \newline Published Date}%
  \end{minipage}
\end{tabular}}
}}
}
%%%%%%%%%% BLOCK: Copyright information
\endgroup

%\linenumbers

\vspace{10pt}
\noindent\rule{\textwidth}{1pt}
\tableofcontents
\noindent\rule{\textwidth}{1pt}
\vspace{10pt}

\section{Introduction}
\label{sec:intro}

A formulation of perturbative gauge theories in terms of dynamical
gauge-invariant variables can make the structure of physical amplitudes more
transparent. Gauge symmetry is then encoded in the geometry of the field
space, so that relations among amplitudes can be read off directly and not just via cancellations
between gauge-dependent contributions. In spontaneously broken theories,
gauge-invariant composite fields offer a concrete realization of this idea.
They can also organize amplitudes into invariant sectors, classified by the number of gauge-invariant fields propagating inside the loops, each of them separately fulfilling  the Slavnov--Taylor (ST) identity of the model~\cite{Quadri2024,Quadri:2026zwv}.

For massless gauge theories, a basic difficulty is that extracting an
invariant representative from the gauge potential usually involves inverse
differential operators. The question is therefore whether the description
can be made local by enlarging the field space. The free gauge complex
provides a useful setting in which to identify precisely where the
non-locality enters and how auxiliary variables can remove it from the
explicit contraction maps.

The strategy discussed here has three steps. First, one contracts the
linearized gauge complex onto invariant representatives. Second, one
replaces the non-local gauge parameter, extracted by the homotopy, with an
independent field obeying a local constraint. Finally, one seeks a BRST
completion encoding the extended field content.

This construction is presented in the case of an Abelian gauge field.
Extension to the non-Abelian case,
where the gauge fields are self-interacting, requires to transfer the interacting gauge structure to the reduced complex through \Linfty\ homotopy transfer.

In this paper we will briefly outline the construction, leaving the complete analysis of the interacting quantum theory to a separate work.

\section{The free gauge complex and its contraction}
\label{sec:free}

Let us consider an Abelian gauge field
$A_\mu$ around the trivial background. 
In flat spacetime, with
$\Box=\partial^\mu\partial_\mu$, the gauge transformation is
\begin{equation}
  \delta_\Lambda A_\mu=\partial_\mu\Lambda.
  \label{eq:gauge}
\end{equation}
The classical action is
\begin{equation}
S=-\frac{1}{4} \int \dd^4x \, 
F^{\mu\nu}F_{\mu\nu} \, , 
\end{equation}
where $F_{\mu\nu}$ is the field strength
$F_{\mu\nu} = \partial_\mu A_\nu - \partial_\nu A_\mu$.
The equation of motion (e.o.m.) reads
\begin{align}
    E_\mu = \frac{\delta S}{\delta A^\mu} = \frac{1}{2} (\square A_\mu - \partial_\mu (\partial A)) \, .
\end{align}

Gauge parameters, fields, e.o.m. and Noether identities belong
to the chain complex
\begin{equation}
  X_1\xrightarrow{\partial_1}X_0
  \xrightarrow{\partial_0}X_{-1}
  \xrightarrow{\partial_{-1}}X_{-2},
  \label{eq:complex}
\end{equation}
with representatives $\Lambda$, $A_\mu$, $E_\mu$ and $f$, respectively. The
differential has degree $-1$ and acts as
\begin{align}
  (\partial_1\Lambda)_\mu&=\partial_\mu\Lambda,\nonumber\\
  (\partial_0 A)_\mu&=\Box A_\mu-\partial_\mu(\partial A),\nonumber\\
  \partial_{-1}E_\mu&=\partial^\mu E_\mu.
  \label{eq:differentials}
\end{align}
The identities
\begin{equation}
  \partial_0\partial_1=0,\qquad
  \partial_{-1}\partial_0=0
  \label{eq:nilpotency}
\end{equation}
express Abelian gauge invariance and the Abelian Noether identity
(the divergence identity for the gauge-field equations). The free
equations are $\partial_0 A_\mu=0$, while gauge-equivalent configurations differ
by an element of $\im\partial_1$.

\subsection{Gauge-invariant representatives}

Choose a formal Green operator $G=\Boxinv$ and write
\begin{equation}
  \psi=G\divA,\qquad
  \widetilde A_\mu=A_\mu-\partial_\mu G\partial A.
  \label{eq:transverse}
\end{equation}
For the contraction identities below, $G$ is assumed to commute with
derivatives and to satisfy $\Box G=G\Box=\id$ (the identity operator) on the chosen domain.

We will limit ourselves to 
perturbation theory around the trivial background, so we will discard
homogeneous solutions and zero modes 
of the G operator. 

Under these assumptions $\delta_\Lambda\psi=\Lambda$ and
$\delta_\Lambda\widetilde A_\mu=0$. Moreover,
\begin{equation}
  \partial^\mu\widetilde A_\mu=0,\qquad
  (\partial_0 A)_\mu=\Box\widetilde A_\mu.
  \label{eq:transverse-eom}
\end{equation}
The representative $\widetilde A_\mu$ is thus invariant but non-local. The non-locality is
associated with the extraction of the gauge parameter $\psi$, rather than
with gauge invariance itself.

The reduced complex contains only transverse fields and transverse equation
variables:
\begin{equation}
  0\longrightarrow\overline X_0
  \xrightarrow{\overline\partial_0}\overline X_{-1}
  \longrightarrow0,
  \qquad \overline\partial_0\widetilde A_\mu=\Box\widetilde A_\mu.
  \label{eq:reduced}
\end{equation}
Thus the reduction removes gauge redundancy, but the reduced off-shell field
space should not be identified with the space of on-shell physical
polarizations before imposing the relevant equations of motion.

In Ref.~\cite{Chiaffrino2021} this construction has been presented with reference to the operator $G$ chosen to be the spatial Laplacian 
$\Delta = \partial^i \partial_i$
after performing the space/time split of the gauge field as $A_\mu=(A_0, A_i)$.
By the Helmholtz theorem $A_i$ is decomposed into the sum of a divergenceless vector and the gradient of a scalar
\begin{equation}
    A_i = \widehat A_i + \partial_i \psi \, .
\end{equation}
The decompostion exists whenever $\Delta$ is invertible, with
\begin{align}
    \psi = \Delta^{-1} (\partial_i A^i) \, .
\end{align}
Then one finds
\begin{align}
\delta_\Lambda A_0 = \dot{\Lambda} \, , \qquad \delta \widehat{A}_i = 0 \, , \qquad \delta \Lambda = 0 \, .
\end{align}
Time locality is preserved but at the price of loosing manifest Lorentz covariance of the formulation. In the present paper we keep the operator $G=\square^{-1}$ and will show how to define a local version of the theory, suited to be treated in the standard Lorentz-covariant perturbation theory. 

\subsection{Projection, inclusion and homotopy}

Let $p:X\to\overline X$ be the projection and
$i:\overline X\to X$ the inclusion operators. 
We will look for a homotopy operator
$h:X_k\to X_{k+1}$,
acting on the graded vector spaces $X_k$, 
such that the contraction identities hold
\begin{equation}
  pi=\id_{\overline X},\qquad
  \id_X-ip=\partial h+h\partial.
  \label{eq:contraction}
\end{equation}
Let us define the transverse operator $P_T$ by 
\begin{align}
(P_TV)_\mu=V_\mu-\partial_\mu G\partial^\nu V_\nu \, .
\end{align}
$P_T$ removes the longitudinal  part of the vector $V_\mu$,
leaving a divergence-free vector. Assuming $G$ is an inverse of
$\Box=\partial^\mu\partial_\mu$ on the chosen function space,
\[
\partial^\mu(P_TV)_\mu
=\partial^\mu V_\mu-\Box G\,\partial^\nu V_\nu=0.
\]
Consequently, $P_T^2=P_T$, as required for a projector.

The non-trivial projections can be written using the transverse operator,
since
\begin{equation}
  p_0(A_\mu)=(P_TA)_\mu,\qquad p_{-1}(E_\mu)=(P_TE)_\mu \equiv \widetilde{E}_\mu \, .
  \label{eq:projection}
\end{equation}
Here $E_\mu$ is an arbitrary element of $X_{-1}$. When it is evaluated on
$E_\mu=\partial_0 A_\mu$, it is already transverse, and
$p_{-1}(\partial_0 A_\mu)=\Box\widetilde A_\mu$.
The inclusions are the natural embeddings of transverse vectors:
\begin{align}
  i_0:\overline X_0\hookrightarrow X_0,
  &\qquad (i_0\widetilde A)_\mu=\widetilde A_\mu,
  &\partial^\mu\widetilde A_\mu=0,
  \\
  i_{-1}:\overline X_{-1}\hookrightarrow X_{-1},
  &\qquad (i_{-1}\widetilde E)_\mu=\widetilde E_\mu,
  &\partial^\mu\widetilde E_\mu=0.
\end{align}
Since $\overline X_1=\overline X_{-2}=\{0\}$, the remaining
inclusions are the zero maps,
\[
  i_1:\{0\}\to X_1,
  \qquad
  i_{-2}:\{0\}\to X_{-2}.
\]
In particular,
\[
  p_0i_0=\id_{\overline X_0},
  \qquad
  p_{-1}i_{-1}=\id_{\overline X_{-1}},
\]
because $P_T$ acts as the identity on transverse vectors.

Moreover
$p_1=p_{-2}=0$.
The homotopy is
\begin{equation}
  h_0(A_\mu)=G\partial A,\qquad h_{-1}(E)=0,\qquad
  (h_{-2}f)_\mu=\partial_\mu Gf.
  \label{eq:homotopy}
\end{equation}
For example, for the vector field,
\begin{equation}
  ((\id-ip)A)_\mu=\partial_\mu G\divA
  =(\partial_1h_0A)_\mu,
\end{equation}
while at the equation of motion degree,
\begin{equation}
  ((\id-ip)E)_\mu=\partial_\mu G\divE
  =(h_{-2}\partial_{-1}E)_\mu.
\end{equation}
At the two ends, $h_0\partial_1\Lambda=\Lambda$ and
$\partial_{-1}h_{-2}f=f$. These relations verify
Eq.~\eqref{eq:contraction} degree by degree. A spatial splitting, such as
one adapted to Coulomb gauge, can preserve locality in time while retaining
spatial inverse operators~\cite{Chiaffrino2021}.

\section{Localizing the field-level contraction}
\label{sec:local}

We now promote $\psi$ to an independent scalar and impose the local differential
constraint
\begin{equation}
  C(A_\mu,\psi)=\Box\psi-\divA=0.
  \label{eq:constraint}
\end{equation}
We define the extended gauge transformation and the invariant variable by
\begin{equation}
  \delta_\Lambda(A_\mu,\psi)=(\partial_\mu\Lambda,\Lambda),
  \qquad B_\mu=A_\mu-\partial_\mu\psi.
  \label{eq:B}
\end{equation}
Both $B_\mu$ and $C$ are invariant under the Abelian gauge transformation. In
coordinates $(B_\mu,\psi)$, the gauge differential simply acts as
\begin{equation}
  \delta_\Lambda(B_\mu,\psi)=(0,\Lambda).
\end{equation}
The gauge parameter and the scalar therefore form a contractible two-term
sector $K$, with differential $\Lambda\mapsto\psi=\Lambda$.
At the level of gauge parameters and fields, the enlarged space splits into
an invariant sector and $K$, as can be seen as follows.

% Lo spazio X_ext viene completato con il settore del vincolo nella
% formulazione dinamica; qui sono esplicitati i soli parametri e campi.
%%%
Let $\mathcal S$ and $\mathcal V$ denote the spaces of scalar
and vector fields, respectively, and define the e.o.m. operator by
\begin{equation}
  (\mathcal E A)_\mu
  =\Box A_\mu-\partial_\mu\partial^\nu A_\nu.
\end{equation}
We complete the extended parameter--field sector to the complex
\begin{equation}
  X^{\mathrm{ext}}_1
  \xrightarrow{\partial^{\mathrm{ext}}_1}
  X^{\mathrm{ext}}_0
  \xrightarrow{\partial^{\mathrm{ext}}_0}
  X^{\mathrm{ext}}_{-1}
  \xrightarrow{\partial^{\mathrm{ext}}_{-1}}
  X^{\mathrm{ext}}_{-2},
\end{equation}
where
\begin{equation}
  X^{\mathrm{ext}}_1=\mathcal S,
  \qquad
  X^{\mathrm{ext}}_0=\mathcal V\oplus\mathcal S,
  \qquad
  X^{\mathrm{ext}}_{-1}=\mathcal V\oplus\mathcal S,
  \qquad
  X^{\mathrm{ext}}_{-2}=\mathcal S.
\end{equation}
An element of the e.o.m. space $X^{\mathrm{ext}}_{-1}$ is denoted by $(E_\mu,c)$,
where $c$ represents the constraint equation. The differentials are
\begin{align}
  \partial^{\mathrm{ext}}_1\Lambda
  &=(\partial_\mu\Lambda,\Lambda),
  \\
  \partial^{\mathrm{ext}}_0(A_\mu,\psi)
  &=\bigl(\mathcal E A_\mu,\Box\psi-\partial^\mu A_\mu\bigr),
  \\
  \partial^{\mathrm{ext}}_{-1}(E,c)
  &=\partial^\mu E_\mu.
\end{align}
They satisfy
\begin{equation}
  \partial^{\mathrm{ext}}_0
  \partial^{\mathrm{ext}}_1=0,
  \qquad
  \partial^{\mathrm{ext}}_{-1}
  \partial^{\mathrm{ext}}_0=0,
\end{equation}
by gauge invariance of the e.o.m. operator ${\mathcal E}$ and the identity
$\partial^\mu(\mathcal E A)_\mu=0$.

In the local coordinates
\begin{equation}
  B_\mu=A_\mu-\partial_\mu\psi,
  \qquad
  A_\mu=B_\mu+\partial_\mu\psi,
\end{equation}
the differentials become
\begin{align}
  \partial^{\mathrm{ext}}_1\Lambda
  &=(0,\Lambda),
  \\
  \partial^{\mathrm{ext}}_0(B,\psi)
  &=\bigl(\mathcal E B,-\partial^\mu B_\mu\bigr),
  \\
  \partial^{\mathrm{ext}}_{-1}(E,c)
  &=\partial^\mu E_\mu.
\end{align}
Thus the complex splits as
\begin{equation}
  X^{\mathrm{ext}}\cong\overline X^{\mathrm{loc}}\oplus K,
\end{equation}
where $K$ is the contractible two-term complex
\begin{equation}
  K_1=\mathcal S
  \xrightarrow{\;\id\;}
  K_0=\mathcal S,
\end{equation}
formed by the gauge parameter and the scalar $\psi$.

The retained complex is
\begin{equation}
  \overline X^{\mathrm{loc}}_1=0,
  \qquad
  \overline X^{\mathrm{loc}}_0=\mathcal V,
  \qquad
  \overline X^{\mathrm{loc}}_{-1}
  =\mathcal V\oplus\mathcal S,
  \qquad
  \overline X^{\mathrm{loc}}_{-2}=\mathcal S,
\end{equation}
with differential
\begin{equation}
  \overline\partial_0 B
  =\bigl(\mathcal E B,-\partial^\mu B_\mu\bigr),
  \qquad
  \overline\partial_{-1}(E,c)=\partial^\mu E_\mu.
\end{equation}

Returning to the original coordinates $(A,\psi)$, we define the
projection and inclusion by
\begin{align}
  p^{\mathrm{loc}}_0(A_\mu,\psi)
  &=A_\mu-\partial_\mu\psi,
  &
  i^{\mathrm{loc}}_0(B_\mu)
  &=(B_\mu,0),
  \\
  p^{\mathrm{loc}}_{-1}(E_\mu,c)
  &=(E_\mu,c),
  &
  i^{\mathrm{loc}}_{-1}(E_\mu,c)
  &=(E_\mu,c),
  \\
  p^{\mathrm{loc}}_{-2}f
  &=f,
  &
  i^{\mathrm{loc}}_{-2}f
  &=f.
\end{align}
At degree one, $p^{\mathrm{loc}}_1$ and
$i^{\mathrm{loc}}_1$ are the zero maps.
These are chain maps since
\begin{equation}
  p^{\mathrm{loc}}\partial^{\mathrm{ext}}
  =\overline\partial p^{\mathrm{loc}},
  \qquad
  \partial^{\mathrm{ext}}i^{\mathrm{loc}}
  =i^{\mathrm{loc}}\overline\partial.
\end{equation}

The only nonzero component of the contracting homotopy is
\begin{equation}
  h^{\mathrm{loc}}_0(A_\mu,\psi)=\psi,
  \qquad
  h^{\mathrm{loc}}_{-1}=h^{\mathrm{loc}}_{-2}=0.
\end{equation}
At field degree,
\begin{equation}
  (\id-i^{\mathrm{loc}}p^{\mathrm{loc}})(A_\mu,\psi)
  =(\partial_\mu\psi,\psi)
  =\partial^{\mathrm{ext}}_1
    h^{\mathrm{loc}}_0(A_\mu,\psi),
\end{equation}
while at parameter degree,
\begin{equation}
  h^{\mathrm{loc}}_0
  \partial^{\mathrm{ext}}_1\Lambda=\Lambda.
\end{equation}
At e.o.m. and Noether identity degrees, $i^{\mathrm{loc}}p^{\mathrm{loc}}$
is the identity and both homotopy terms vanish. Consequently,
\begin{equation}
  p^{\mathrm{loc}}i^{\mathrm{loc}}
  =\id_{\overline X^{\mathrm{loc}}},
  \qquad
  \id_{X^{\mathrm{ext}}}
  -i^{\mathrm{loc}}p^{\mathrm{loc}}
  =\partial^{\mathrm{ext}}h^{\mathrm{loc}}
   +h^{\mathrm{loc}}\partial^{\mathrm{ext}}.
\end{equation}
The side conditions also hold:
\begin{equation}
  (h^{\mathrm{loc}})^2=0,
  \qquad
  h^{\mathrm{loc}}i^{\mathrm{loc}}=0,
  \qquad
  p^{\mathrm{loc}}h^{\mathrm{loc}}=0.
\end{equation}
%%%

The corresponding maps are local:
\begin{equation}
  p_0^{\mathrm{loc}}(A_\mu,\psi)=B_\mu,\qquad
  i_0^{\mathrm{loc}}(B_\mu)=(B_\mu,0),\qquad
  h_0^{\mathrm{loc}}(A_\mu,\psi)=\psi.
  \label{eq:localmaps}
\end{equation}
Indeed,
\begin{equation}
  (A_\mu,\psi)-i_0^{\mathrm{loc}}p_0^{\mathrm{loc}}(A_\mu,\psi)
  =(\partial_\mu\psi,\psi)
  =\partial_1^{\mathrm{ext}}h_0^{\mathrm{loc}}(A_\mu,\psi),
  \label{eq:localidentity}
\end{equation}
where $\partial_1^{\mathrm{ext}}\Lambda=(\partial_\mu\Lambda,\Lambda)$.
The local inverse change of coordinates is
$A_\mu=B_\mu+\partial_\mu\psi$.

The constraint plays an essential role in relating this enlarged space to
the original reduction. In the new coordinates,
\begin{equation}
  C(A_\mu,\psi)=-\divB.
  \label{eq:constraintB}
\end{equation}
Thus the constraint enforces the transversality of $B$. On its surface, the
inclusion $(B_\mu,0)$ also satisfies $C=0$; away from it,
Eq.~\eqref{eq:localidentity} is a kinematic contraction identity on the
unconstrained parameter--field sector.

%A contraction of the full dynamical
%complex additionally requires the equation for the constraint and its
%associated identities to be included consistently.

Solving Eq.~\eqref{eq:constraint} with the same prescription used above
gives $\psi=G\divA$ and $B_\mu=\widetilde A_\mu$. Consequently, locality is achieved
in the enlarged description: eliminating the extra variable one recovers the
non-local expression in the original variables.

% Comparison with the FMS formalism

\section{Auxiliary fields and the BRST formulation}
\label{sec:brst}

At the action level, the constraint can be implemented by a Lagrange
multiplier, denoted by $X$ :
\begin{equation}
  S_{\mathrm{aux}}=\int\dd^4x\, X(\Box\psi-\divA).
  \label{eq:Saux}
\end{equation}
Its equation of motion is
\begin{equation}
  \frac{\delta S_{\mathrm{aux}}}{\delta X}=C(A_\mu,\psi)=0.
\end{equation}
Formally, integration over $X$ imposes
at the path-integral level $\delta[C]$. Subsequent elimination
of $\psi$ reproduces the transverse representative, up to the determinant
of the constraint operator (and neglecting 
non-trivial homogeneous solutions), yielding a factor
$\sim 1/| {\rm det} ~ \square|$ that is $A_\mu$-independent and therefore can be dropped by the 
path-integral normalization factor.

We notice that gauge invariance of $S_{\mathrm{aux}}$
 is not by itself a proof that the enlarged
theory has the same physical cohomology as the original one. For that purpose we need a set of additional constraint ghosts and antighosts.

The construction works as follows. 
In the Abelian case, the gauge BRST differential acts as
\begin{equation}
  sA_\mu=\partial_\mu \omega,\qquad s\psi=\omega,\qquad s\omega=0,
  \qquad sB_\mu=0,
  \label{eq:BRST}
\end{equation}
where $\omega$ is the gauge ghost. 
We now introduce an extra $\sc$ BRST differential 
$\mathfrak{s}$ so that
\begin{align}
    \mathfrak{s} \bar c = C(A_\mu,\psi) \, , \quad
    \mathfrak{s} X = c \, \quad
    \mathfrak{s} c = 0 \, , 
    \label{eq:BRST.constr}
\end{align}
while $\mathfrak{s}$ is zero on other fields.
Then we promote $S_{\mathrm{aux}}$ to
\begin{equation}
    S_{\mathrm{aux+constr}} = \mathfrak{s} \int d^D x \,
    \Big ( \bar c X  \Big ) = 
    \int \dd^4 x \Big ( X (\Box\psi-\divA) -
    \bar c c 
    \Big ) \, .
\end{equation}
$S_{\mathrm{aux+constr}}$ is both gauge BRST-invariant, since $sC=0$,  and $\mathfrak{s}$-invariant.

The relevant algebraic mechanism is the triviality of contractible pairs.
For a BRST differential $q$ we say that the independent variables $u,v$ form a contractible pair (also known as a BRST doublet) if they satisfy
\begin{equation}
  qu=v,\qquad qv=0 \, .
  \label{eq:doublet}
\end{equation}
We see from Eqs.(\ref{eq:BRST})
 and (\ref{eq:BRST.constr}) that 
 both $(\psi, \omega)$ and $(X,c)$ are BRST doublets~\cite{Barnich:2000zw,Quadri:2002nh}, so they do not alter the physical cohomology of the theory.

Introducing the compensating scalar $\psi$ has an important consequence
for the space of gauge-invariant operators. Under
$\delta_\Lambda A_\mu=\partial_\mu\Lambda$ and
$\delta_\Lambda\psi=\Lambda$, the combination $B_\mu(A,\psi)$ is gauge-invariant.
Thus the local
mass term
\begin{equation}
  \frac{m^2}{2} B^2 
\end{equation}is admissible in the enlarged formulation. This is the Abelian
St\"uckelberg mechanism for introducing a gauge-invariant
mass~\cite{Stueckelberg:1938hvi,Ferrari:2004pd}.

This observation does not contradict the usual characterization
of local gauge-invariant operators in the pure massless Abelian
theory. In the polynomial algebra generated by $A_\mu$ and its
derivatives, strictly gauge-invariant operators are generated by
the field strength and its derivatives. 
$B_\mu$ belongs to a different local field algebra, which
also contains the independent scalar $\psi$.

The relation between the two descriptions becomes explicit upon
eliminating $\psi$. Solving its differential constraint with the
chosen Green-operator prescription gives
\begin{equation}
  \psi=G\partial^\nu A_\nu,
  \qquad
  B_\mu
  =A_\mu-\partial_\mu G\partial^\nu A_\nu,
\end{equation}
as in Eq.~\eqref{eq:transverse}. The same mass term is therefore
nonlocal when expressed solely in terms of $A_\mu$. More
generally, polynomials in $B_\mu$ and finitely many of its
derivatives define local gauge-invariant operators in the
enlarged formulation and corresponding, generally nonlocal,
operators in the original variables. The resulting invariant
algebra is understood subject to the transversality constraint
when that constraint is imposed. The enlarged formulation thus
provides local representatives of this specified algebra of
nonlocal gauge-invariant operators.

\section{Interactions and homotopy transfer}
\label{sec:interactions}

In this section we outline how the preceding construction can be
extended to non-Abelian gauge theories. We focus on the classical
algebraic framework and the conditions required for a local
description. The full BRST formulation and its quantum analysis
will be presented separately.

The starting point is the free gauge complex, which, as discussed in Section~\ref{sec:free}, contains gauge
parameters, fields, e.o.m. and Noether identities. In an
interacting theory, the differential $\partial$ is supplemented
by multilinear brackets encoding the nonlinear terms. Together,
these maps define an \Linfty\ algebra on the graded vector space
$X$.

We use the shifted homological convention, in which all brackets
are graded symmetric and have degree $-1$:
\begin{equation}
  \ell_n:S^nX\longrightarrow X,
  \qquad
  \deg(\ell_n)=-1,
  \qquad
  \ell_1=\partial.
  \label{eq:brackets}
\end{equation}
Here $S^nX$ denotes the graded symmetric tensor power of $X$.
Thus, for homogeneous elements, exchanging adjacent inputs gives
the sign $(-1)^{|x_i||x_{i+1}|}$. This convention is related by
suspension to the usual formulation with graded antisymmetric
brackets. We work perturbatively around a classical solution,
so that no zero-ary bracket is present.

For a field $a\in X_0$ and a gauge parameter $\Lambda\in X_1$,
the equations and gauge transformations take the form
\begin{align}
  \mathcal F(a)
  &=
  \sum_{n\geq1}\frac{1}{n!}
  \ell_n(a,\ldots,a),
  \nonumber\\
  \delta_\Lambda a
  &=
  \sum_{n\geq0}\frac{1}{n!}
  \ell_{n+1}(\Lambda,a,\ldots,a).
  \label{eq:interacting}
\end{align}
The equations of motion are $\mathcal F(a)=0$. At linear order,
these expressions reduce to
$\mathcal F(a)=\partial_0a$ and
$\delta_\Lambda a=\partial_1\Lambda$.
The higher brackets describe interactions, field-dependent gauge
transformations and the corresponding nonlinear identities.

These compatibility conditions can be expressed compactly using
the coaugmented graded symmetric coalgebra
\begin{equation}
  S^c(X)=\bigoplus_{n\geq0}S^nX.
\end{equation}
Each bracket $\ell_n$ extends uniquely to a coderivation
$\widehat\ell_n$ on this coalgebra. The \Linfty\ structure is
encoded by
\begin{equation}
  Q=\sum_{n\geq1}\widehat\ell_n,
  \qquad
  \deg(Q)=-1,
  \qquad
  Q(1)=0,
  \qquad
  Q^2=0.
  \label{eq:coderivation}
\end{equation}
The component of $Q^2=0$ with one input gives $\partial^2=0$.
The components with more inputs express the compatibility of the
differential with the interactions and gauge transformations.
In particular, they imply the nonlinear Noether identity
\begin{equation}
  \sum_{n\geq0}\frac{1}{n!}
  \ell_{n+1}\bigl(\mathcal F(a),a,\ldots,a\bigr)=0,
\end{equation}
whose linear part is
$\partial_{-1}\partial_0a=0$.

Homotopy transfer starts from a contraction of the complete free
complex onto a reduced complex $(\overline X,\overline\partial)$.
The projection $p$ and inclusion $i$ must be chain maps, and the
degree-$+1$ homotopy $h$ must satisfy
\begin{equation}
  pi=\id_{\overline X},
  \qquad
  \id_X-ip=\partial h+h\partial.
\end{equation}
These relations must hold at every degree, including the e.o.m.
and identity sectors. A contraction defined only on gauge
parameters and fields is  insufficient to transfer the
full interacting theory.

Given these data, the homotopy transfer theorem equips
$\overline X$ with brackets $\overline\ell_n$ satisfying the
\Linfty\ identities. The first two brackets are
\begin{equation}
  \overline\ell_1=p\ell_1i=\overline\partial,
  \qquad
  \overline\ell_2(x_1,x_2)
  =p\ell_2(ix_1,ix_2).
  \label{eq:transferred}
\end{equation}
The higher brackets are sums of tree compositions: the original
brackets occur at the vertices, the homotopy $h$ on internal
edges, the inclusion $i$ on the inputs and the projection $p$
on the output. The graded signs and symmetry factors are fixed
by the chosen convention.

The projection and inclusion must also be extended beyond their
linear parts. On fields, their perturbative expansions have the
form
\begin{align}
  \mathcal P(a)
  &=pa+\frac{1}{2}\mathcal P_2(a,a)+\cdots,
  \nonumber\\
  \mathcal I(\overline a)
  &=i\overline a
    +\frac{1}{2}\mathcal I_2(\overline a,\overline a)+\cdots.
\end{align}
These corrected maps relate the original and transferred
equations and gauge structures. For a reduction to invariant
variables, the corrected projection must map the original gauge
transformations to trivial transformations of the reduced fields.
This is the mechanism underlying the construction of
gauge-invariant variables by homotopy transfer discussed in
Ref.~\cite{Chiaffrino2021}.

Locality depends on the specific brackets and contraction maps
used in this construction. At each fixed perturbative order,
the transfer involves finitely many compositions. If the
original brackets and the maps $p$, $i$ and $h$ are local
differential maps, these compositions remain local. Conversely,
inverse differential operators appearing in the contraction
will generally enter the transferred brackets and maps.

For the enlarged formulation of Sect.~\ref{sec:local}, the
relevant input is therefore a local contraction of the entire
extended complex, together with a compatible nonlinear
completion. The constraint equations and their identities must
be included in this completion. Locality in the enlarged
variables does not imply locality after the auxiliary variables
have been eliminated: solving their differential constraints
may reintroduce the inverse operators that occur in the
original description.

The compensating scalar also requires a nonlinear completion.
For Yang--Mills theory,
\begin{equation}
  \delta_\Lambda A_\mu
  =\partial_\mu\Lambda+g[A_\mu,\Lambda].
\end{equation}
The Abelian expressions
$\delta_\Lambda\psi=\Lambda$ and
$B_\mu=A_\mu-\partial_\mu\psi$
cancel only the linearized gauge variation. In the interacting
theory they must be replaced by expressions of the form
\begin{equation}
  \delta_\Lambda\psi=\Lambda+O(g),
  \qquad
  B_\mu=A_\mu-\partial_\mu\psi+O(g),
\end{equation}
with the corrections chosen so that $B_\mu$ is invariant under
the full gauge transformation. The extended constraint must
likewise be completed consistently with this symmetry.

Finally, homotopy transfer establishes a classical algebraic
relation between the two descriptions. Transferring an action
requires additional structure:  an appropriate Batalin-Vilkovisky formulation, together with transfer maps
that respect it. Quantum equivalence further requires control
of  renormalization and proof that the Slavnov-Taylor identities still hold at the quantum level. The classical
identity $Q^2=0$ is a necessary consistency condition, but
does not by itself establish equivalence to all orders in
the quantum theory.

\section{Relation to the FMS construction}
\label{sec:FMS}

In the Higgs phase, the Fr\"ohlich--Morchio--Strocchi (FMS) construction
relates invariant composite operators to elementary fields through an
expansion about a non-zero scalar background~\cite{FMS1981}. For the Abelian case, let
\begin{equation}
  \phi=\frac{1}{\sqrt2}(v+\sigma+i\chi),\qquad
  D_\mu=\partial_\mu-ieA_\mu.
\end{equation}
The invariant vector composite field is
\begin{align}
  a_\mu
  &=\frac{i}{ev^2}\left[
     2\phi^\dagger D_\mu\phi-\partial_\mu(\phi^\dagger\phi)
     \right]\nonumber\\
  &=\frac{1}{ev^2}\left\{
    e\big[(v+\sigma)^2+\chi^2\big]A_\mu
    -(v+\sigma)\partial_\mu\chi+\chi\partial_\mu\sigma
    \right\}\nonumber\\
  &=A_\mu-\frac{1}{ev}\partial_\mu\chi
    +\text{terms at least quadratic in the fluctuations}.
  \label{eq:FMS}
\end{align}
At leading order, $\chi/(ev)$ plays the role of a compensating scalar.
The expansion and its normalization require $v\neq0$; a perturbative
coordinate change relies on a non-vanishing scalar Higgs background vacuum expectation values. 
The
gauge-invariant composite describes the physical channel, while elementary
fields provide useful coordinates for perturbative calculations and LSZ
reduction. A correspondence between elementary and invariant spectra
depends on the gauge group and scalar sector~\cite{Quadri2024}.

The two constructions share the use of a scalar compensator, but obtain it
in different ways. Their main features are compared in
Table~\ref{tab:comparison}.
\begin{table}[htbp]
  \centering
  \small
  \renewcommand{\arraystretch}{1.18}
  \begin{tabularx}{\linewidth}{@{}lXX@{}}
    \toprule
    & FMS expansion & Enlarged-complex localization \\
    \midrule
    Invariant variables & Local composites of the Higgs and gauge fields
      & Local expressions involving auxiliary fields \\
    Scalar input & Non-zero background $v$
      & Independent compensator and differential constraint \\
    Geometric mechanism & Expansion around the Higgs background
      & Contractible gauge sector \\
    Regime discussed & Higgs phase
      & Perturbative interacting extension;
      St\"uckelberg mass term\\
    \bottomrule
  \end{tabularx}
  \caption{Comparison of the mechanisms underlying the two descriptions.}
  \label{tab:comparison}
\end{table}

Thus, the enlarged-complex approach separates invariant and
redundant variables and allows a gauge-invariant St\"uckelberg
mass term,
$\frac{m^2}{2}(A_\mu-\partial_\mu\psi)^2$,
to be introduced without a Higgs field. The additional constraint
and its BRST implementation maintain the correspondence with
the original massless theory when this mass deformation is
absent; including it yields a gauge-invariant massive extension.

\section{Conclusion and outlook}
\label{sec:outlook}

The free gauge complex makes the origin of the standard non-local invariant
representative explicit: it lies in the homotopy $h_0(A)=\Boxinv\divA$.
Introducing an independent scalar replaces this map by the local
$h_0^{\mathrm{loc}}(A,\psi)=\psi$, while a differential constraint recovers
the original representative on elimination of the extra field. The
parameter--field sector then separates locally into invariant coordinates
and a contractible gauge sector.

The next steps are the completion of the extended equation complex, the
construction of the auxiliary BRST sector, and the transfer of the nonlinear
brackets. These ingredients determine whether the local formulation
preserves the physical content and remains local at the desired
perturbative order. 
The connection with the St\"uckelberg mechanism in the Abelian theory has been discussed.
The comparison with the FMS mechanism shows how a related invariant
description can be obtained from a Higgs background, under the crucial condition of a non-vanishing vacuum expectation value, while the localization
mechanism itself does not rely on a non-zero vacuum expectation value.

A prospective application is the complete study of the renormalization of the interacting gauge complex in the non-Abelian case, in particular in order to elucidate the connection with the St\"uckelberg non-Abelian mechanism for mass generation.

\section*{Acknowledgements}
The Author thanks the organizers of the
36th International Colloquium on Group Theoretical Methods in Physics
Valladolid for the pleasant and fruitful discussions.

\paragraph{Funding information}
\textit{Financial support by INFN, Sezione di Pisa is gratefully acknowledged.}

\end{document}